\documentclass[sigconf]{acmart}
\AtBeginDocument{%
  \providecommand\BibTeX{{%
    \normalfont B\kern-0.5em{\scshape i\kern-0.25em b}\kern-0.8em\TeX}}}

\usepackage{amsmath}
\DeclareMathOperator*{\argmax}{arg\,max}

\setcopyright{acmcopyright}
\copyrightyear{2021}
\acmYear{2021}
\acmDOI{10.1145/1122445.1122456}

\acmConference[NetEcon '21]{The 16th Workshop on the Economics of Networks, Systems and Computation}{July 23, 2021}{Budapest, Hungary}
\acmBooktitle{The 16th Workshop on the Economics of Networks, Systems and Computation, July 23, 2021, Budapest, Hungary}
\acmPrice{15.00}
\acmISBN{978-1-4503-9999-9/18/06}

\begin{document}

\title{A Phase Transition in Large Network Games}


\author{Abhishek Shende}
\affiliation{%
  \institution{University of Texas, Austin}
  \city{Austin}
  \country{USA}}
\email{ashende@utexas.edu}

\author{Deepanshu Vasal}
\affiliation{%
  \institution{Northwestern University}
  \city{Evanston}
  \country{USA}}
\email{dvasal@umich.edu}

\author{Sriram Vishwanath}
\affiliation{%
  \institution{University of Texas, Austin}
  \city{Austin}
  \country{USA}}
\email{sriram@utexas.edu}

\renewcommand{\shortauthors}{Shende and Vasal, et al.}

\begin{abstract}
In this paper, we use a model of large random network game where the agents plays selfishly and are affected by their neighbors, to explore the conditions under which the Nash equilibrium (NE) of the game is affected by a perturbation in the network. We use a phase transition phenomenon observed in finite rank deformations of large random matrices, to study how the NE changes on crossing critical threshold points. Our main contribution is as follows: when
the perturbation strength is greater than a critical point, it impacts the NE of the game, whereas when this perturbation is below this critical point, the NE remains independent of the perturbation parameter. This demonstrates a phase transition in NE which alludes that perturbations can affect the behavior of the society only if their strength is above a critical threshold. We provide numerical examples for this result and present scenarios under which  this phenomenon could potentially occur in real world applications.    
\end{abstract}


\maketitle

\section{Introduction}
The ever increasing interactions between people in the world has motivated the analysis of network data in a range of disciplines and applications, appearing in such diverse areas as commerce, sociology, epidemiology, computer science, and national security. Network data is characterized by the edges between nodes and edge weights,through an adjacency matrix, wherein the action of individual player is affected by the actions of its neighbors. These actions not only affect the individuals but also the overall society. These outcomes can be modeled and studied through the framework of network games, focusing on understanding the effect of properties of network on Nash equilibria (NE). The authors in \cite{Parise2017a} and \cite{Jackson2015} study the properties of Nash equilibrium in different types of network games and how these are impacted by network structure.

A majority of  real-world networks are large networks, with numerous players and interactions. Many random matrix models have been created that incorporate features of real-world systems, such as, network composition, flexibility, and recurrent motifs~\cite{Newman2018,Jackson2010}. These network models appear in the study of complex systems, such as social networks, economic markets, signal processing and natural ecosystems. Characterizing the conditions on such networks for Nash equilibrium, even approximately, helps in  understanding the results when players play selfishly. For large network games, we assume that the adjacency matrix is a random matrix, generated through a given distribution. 

In statistical physics, the Ising model on large random graphs helps to study the phase transition phenomenon observed in the real world. It is shown in \cite{Landau1976Finite-sizeLattice,Stanley1999ScalingPhenomena}, that the critical point behavior occurs for ferromagnetic interactions, where the magnetization vanishes as the continuously increasing temperature crosses a certain threshold.
The phase transition between water and steam is an example of a phase transition occurring at a critical, or Curie temperature. This transition can be modeled with the Ising model, where the nodes are points in space, and the presence or absence of a molecule is generated randomly, like spins. Thus the magnetism corresponds to the density of the H$_2$O. At a hundred degrees Celsius, the density changes substantially; a phase transition. The author Malcolm Gladwell in \cite{Gladwell2002} argues that there exist similar `tipping points' in the society that are critical moments when a minor change makes all the difference. Tipping points are derived from ideas in epidemiology, the study of the spread of viruses and other diseases. The example of a simplified flu epidemic, as provided by Gladwell, describes how the start of Christmas season is a tipping point which leads to exponential increase in transmission rate due to such a simple cause as crowded Christmas shopping and cold weather. The theory of epidemic transmission is applicable to  many ideas, products, messages, and behaviors we find in society can be characterized by their rapid, exponential spread through our population. The resurgence of brands like Hush Puppies and viral growth of new ones like Airwalk is contributed to early adoption and focus of marketing on influencers in the social network which the book calls as connectors, mavens, and salesmen. Through our paper, we try to use a game theoretic notion to mathematically study how tiny shift across the threshold can create huge effects by propagating the cause throughout the network. 

In this paper, we explore a phase transition phenomenon seen in large random matrices in the context of a network game, and observe how the Nash Equilibrium of a game changes on varying a certain parameter. The authors in \cite{Benaych-Georges2009} have studied the extreme eigenvalues and eigenvectors of finite, low rank perturbations of random matrices. A phase transition phenomenon occurs whereby the large matrix limit of the extreme eigenvalues of the perturbed matrix differs from that of the original matrix if and only if the eigenvalues of the perturbing matrix are above a certain critical threshold. Our results focus on the `spiked' random matrix models with Wigner and Wishart random ensemble for additive and multiplicative perturbations respectively. The key to applying the Baik, Ben Arous and Peche - BBP phase transition results from \cite{Baik2004} lies in being able to compute the Cauchy or T transforms of the probability measure and their associated functional inverses. We use results on Wigner studied in \cite{Bassler2008, Capitaine2009,Feral2007,Hoyle2007} and Wishart studied in \cite{Baik2004,Baik2006,Karoui2005,Nadler2008} random ensemble where the transforms and their inverses can be expressed in closed form.
Through previous work in \cite{Parise2017a,Bramoulle2016,Naghizadeh2017} on how network structure affects the properties of Nash Equilibrium, we know that largest eigenvalue/eigenvector of the adjacency matrix plays a major role in determining  the conditions of various properties of Nash Equilibrium of Linear Quadratic (LQ) network games.
In this paper, we analyse perturbations to the adjacency matrix of large network games, focusing on how the changes to structure of network leads to a phase transition phenomenon of the NE of the perturbed network, due to its dependence on the extreme eigenvalue/eigenvector. To the best of our knowledge this is the first paper that makes a connection between random matrix theory and network games to show a phase transition in large social networks.

The application of such property can be experienced in different domains. The social media enables the communication among people, and promotes several group activities in the society. Based on data from platforms like Facebook, Twitter, etc., some users like celebrities, athletes, or politicians have significantly more followers than the rest. In a network graph terms, these users are the nodes that have much more influence than the rest of nodes. These social media influencers have a loyal follower base, achieving a high level of engagement on their content, such as images, trends, videos, etc., heightening their power of persuasion. Using our analysis, we can see that by introducing a perturbation through a 'teleportation' term into the adjacency matrix of a random network over a critical threshold, the beliefs of the participants in the network can depend on properties of the perturbation strength. This change in beliefs can lead to adoption of new technology or products through changing the influence of neighbors. 

Some examples of network games with linear-quadratic model to observe a phase transition are public goods game \cite{Allouch2015,Naghizadeh2018}, influence of peers in education \cite{Calvo-Armengol2009,Ballester2006} and to model criminal social interactions\cite{Calvo-Armengol2004}. Crime and delinquency are related to connections in social networks, where delinquents often have friends who have committed offenses, and social ties are a means of influence to commit crimes. 
The `tipping point' concept of using perturbation to affect the NE suggests that the properties of friendship networks should be taken into account to better understand peer influence on delinquent behavior and to craft delinquency-reducing policies.

Another application of the linear quadratic model is to model collaboration between firms as presented in \cite{Koenig2014}. Collaboration takes a variety of forms which includes creation and sharing of knowledge about markets and technologies, setting market standards and sharing facilities. The effects of peers has been evident in case studies on the adoption of high yielding hybrid crops by the farmers during the Dust Bowl in USA in late-1920s and 1930s \cite{McLeman2014}. The farmers were reluctant and slow to adopt hybrid crops, largely contributing to expensive switch and very few neighbors using hybrid crops. It was the focus by the government on young farmers over older farmers in combination with outreach by the USDA (headed by hybrid corn pioneer Henry A. Wallace), which can be interpreted as external perturbation on the existing network,that convinced Midwestern farmers to adopt the new seed \cite{Meyers2019,Moscona2021}.



\section{Nash Equilibrium in LQ game}
A network game $\mathcal{G}$ with set of $N$ players, is played over a weighted directed network whose structure is captured by an $n\times n$ adjacency matrix $G$. The $(i,j)$th entry of $G$, denoted by $g_{ij}$, represents the strength and type of influence of player $j$'s strategy on the utility function of player $i$. The positive (negative) $g_{ij}$ represent strategic substitutes (complements) where an increase in neighbor $j$'s actions leads to a corresponding decrease (increase) in player $i$'s action. The action of a player $i$ is given by $x_i \in \mathcal{X}_i \subseteq \mathbb{R}_{\geq0}$, and so $x = (x_1,x_2\dots x_n) \in \mathcal{X} = \prod_{i=1}^N \mathcal{X}_i$. 
Each player $i$ $\in$ $\mathbb{N}[1,N]$ chooses their action $x_i \in \mathbb{R}_{\geq0}$ to maximize a utility function:
\[J_i(x_i,z_i(x)),\]
which in turn depends on their own action $x_i$ and on the aggregate  neighbors' strategies $z_i(x)$, defined by the weighted linear combination 
\[z_i(x) = \sum_{j=1,j\neq i}^{N} g_{ij}x_j.\]
The best response for player $i$, i.e., the action that maximizes the utility function is defined as
\[B_i(z_i(x)) := \argmax_{x_i} J_i(x_i,z_i(x)).\]
The set of actions within which no player has an incentive for unilateral deviations (i.e., each player is playing a best response to other player's actions) is a Nash equilibrium. Mathematically, a vector $x^{*}  = (x^*_1, \dots x^*_n)$,  is a Nash equilibrium (NE) if, for all players $i$ $\in$ $\mathbb{N}[1,N]$, $x^*_i$ $\in$ $B_i(z_i(x))$.

A linear quadratic (LQ) network game is one where each agent chooses a scalar strategy $x_i \geq 0$ in order to maximize the linear quadratic utility function:

\begin{equation} \label{eq:1}
J_i(x_i,z_i(x))=  [z_i(x) + a_i]x_i - \frac{1}{2}q_i(x_i)^2
\end{equation}

with $a_i,q_i$ $\in$ $\mathbb{R}$. 

The Nash equilibrium (NE) for an LQ game can be derived from the first-order necessary condition to maximize the utility function for each player $i$ given by:
\begin{equation}\label{eq:2}
\frac{\partial J_i(x_i,z_i(x))}{\partial x_i} =  \sum_{j=1,j\neq i}^N g_{ij}x_j + a_i - q_ix_i  = 0.
\end{equation}
This results in
\begin{equation} \label{eq:3}
    q_ix^*_i = a_i + \sum_{j=1,j\neq i}^N g_{ij}x^*_j.    
\end{equation}

\subsection{Assumption}
To prove our main results, we require assumptions on the linear quadratic utility function in \eqref{eq:1}. In order to proceed, we impose the assumption that the vector $a= (a_1, \dots a_n) = 0 $ and $q_i = \lambda_{max}-g_{ii}$, where $\lambda_{max}$ is the maximum eigenvalue of the adjacency matrix $G$.

Using these assumptions, \eqref{eq:3} in matrix form, is 
\begin{equation} \label{eq:4}
    (\lambda_{max}I-G)x^* = 0,
\end{equation}
  where $I$ is $N\times N$ identity matrix 

\begin{equation} \label{eq:5}
    \lambda_{max}x^* = Gx^*
\end{equation}

Thus, we see that the NE $x^*$ is the eigenvector of $G$ corresponding to the maximum eigenvalue $\lambda_{max}$.

For an eigenvector to be a valid NE strategy, it has to satisfy the condition $x_i \in \mathbb{R}_{\geq0}$ where $x_i$ is each element of the eigenvector. Throughout this paper, we only consider perturbed adjacency matrices that are symmetric and element wise positive, which guarantees that the eigenvector corresponding to maximum eigenvalue has positive entries, through Perron-Frobenius theorem.



\section{Additive Perturbation in LQ games}
Our model for the adjacency matrix of a large network is in the category of the random matrices with fixed-rank deformation, which includes the signal-plus-noise model as typical example. A vast amount of work has been devoted to understanding the limiting behavior of the extreme eigenvalues and the associated eigenvectors of the deformed models. Since the seminal work of authors in \cite{Baik2004}, there has significant study to understand that the extreme eigenvalues undergo a so-called BBP phase transition along with the change of the strength of the deformation. There exists a critical threshold such that the extreme eigenvalue of the perturbed matrix will be within the end points of the spectral distribution if the strength of the deformation is less than or equal to the threshold, and will otherwise be outside of support of the limiting spectral distribution. 

Assume $X$ be an $n \times n$ symmetric Gaussian Wigner matrix, with independent, zero mean normal distribution with variance $\sigma^2/2n$ on off diagonal and $\sigma^2/n$ on diagonal. It has been shown that the spectral measure of $X$ converges almost surely to the well known semi-circle distribution with density 
\begin{equation} \label{eq:6}
    d\mu_x(x) = \frac{\sqrt{4\sigma^2-x^2}}{2\sigma^2\pi}dx \mbox{ for } x\in [-2\sigma,2\sigma].
\end{equation}
Henceforth, $\xrightarrow{\text{a.s.}}$ denotes almost sure convergence. The result from \cite{Anderson2009} shows that the extreme eigenvalues converge almost surely to the endpoints of the support i.e. $\lambda_{max}(X) \xrightarrow{\text{a.s.}} 2\sigma$. 

Let perturbation matrix $P$, be a $n \times n$ symmetric matrix with rank r. Its eigenvalues are $\theta_1 \geq \dots \geq \theta_s >0 >\theta_{s+1} \geq \dots \geq \theta_r$. By phase transition theorem, as $n \rightarrow \infty$ we have, for 

\begin{equation} \label{eq:7}
\tilde{X} = X + P,
\end{equation}

\begin{equation} \label{eq:8}
    \lambda_{max}(\tilde{X}) \xrightarrow{\text{a.s.}} 
    \begin{cases}
    \theta_1 + \frac{\sigma^2}{\theta_1} & \text{if } \theta_1 > \sigma \\
    2\sigma & \text{otherwise}
    \end{cases}
\end{equation}


\begin{figure}[htbp]
\centerline{\includegraphics[width=0.99\columnwidth]{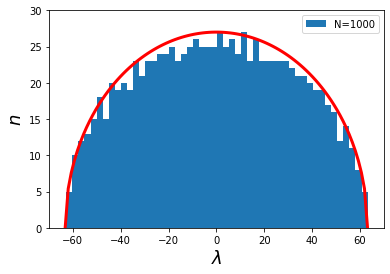}}
\caption{Histogram of eigenvalues of Wigner matrix. The red curve represents the density of the semi-circle distribution}
\label{fig:1}
\end{figure}

\begin{figure}[htbp]
\centerline{\includegraphics[width=0.99\columnwidth]{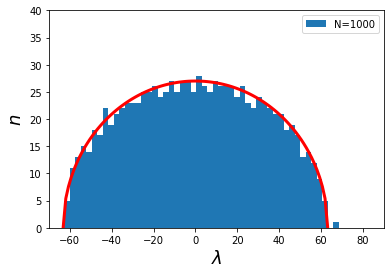}}
\caption{Histogram of eigenvalues of rank 1 perturbation of Wigner matrix. The red curve represents the density of the semi-circle distribution}
\label{fig:2}
\end{figure}
Figure \ref{fig:1} shows the blue histogram of the eigenvalues of a Wigner matrix with $n=1000$, where the red curve follows the semi-circle law from \eqref{eq:6}. 
The extreme eigenvalues are within the boundaries of the semi-circle. The blue histogram of the Figure \ref{fig:2} is that of the eigenvalues of the same Wigner matrix but perturbed this time by a symmetric rank 1 matrix as in \eqref{eq:7}.  The outlier represented here by blue dot outside red curve is exhibited by the extreme eigenvalue of deformed matrix out of the bulk spectrum. This shows the phase transition phenomenon from \eqref{eq:8} when the $\theta > \sigma$

In the setting where $r = 1$ and $P = \theta uu^{T}$, let $\tilde{u}$ be a unit-norm eigenvector of $\tilde{X}$ associated with its largest eigenvalue. The parameter $\theta$ represents the signal to noise ratio. By eigenvector phase transition theorem, for norm of the eigenvector projection we have,
\begin{equation} \label{eq:9}
    |<\tilde{u},u>|^2 \xrightarrow{\text{a.s.}} 
    \begin{cases}
    1 - \frac{\sigma^2}{\theta^2} & \text{if } \theta \geq \sigma \\
    0 & \text{if } \theta < \sigma
    \end{cases}
\end{equation}

\subsection{Transitions in LQ game}
In the context of LQ game, the matrix $X$ is the  $n \times n$ symmetric Gaussian Wigner matrix noise that models the interactions between different participants of a large network. Let $P = \theta uu^T$ so that $\theta$ and $u$ are largest eigenvalue and eigenvector respectively, be the adjacency matrix of the perturbations, leading to deviations in the impact of interactions between different players. 

Considering an additive perturbation as in \eqref{eq:7},the adjacency matrix of the new network is $\tilde{X}$. From \eqref{eq:5}, since the NE of the game described in \eqref{eq:1} is the eigenvector of the adjacency matrix corresponding to the maximum eigenvalue, the NE for game with $\tilde{X}$ is  $\tilde{x}^* = \lambda_{max}(\tilde{X}) = \tilde{u}$. 

From the result of \eqref{eq:9}, for the LQ game, the Nash Equilibrium of the perturbed game is dependent on the parameter $\theta$ of the perturbation. 
A transition phenomenon is observed for values of $\theta$ greater than the threshold value of $\sigma$ when the NE given by $\tilde{u}$ is dependent on the network parameters i.e.$\theta$ and $\sigma$. Below this threshold value, the NE does not depend on the strength of perturbation and cannot be affected as desired by modifying the strength of network connections. 

 

\subsection{Numerical Examples}
\begin{figure}[htbp]
\centerline{\includegraphics[width=0.99\columnwidth]{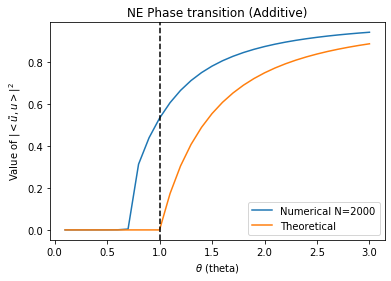}}
\caption{Comparison of transition in NE for additive perturbation of LQ game with $\sigma = 1$.}
\label{fig:3}
\end{figure}

\begin{figure}[htbp]
\centerline{\includegraphics[width=0.99\columnwidth]{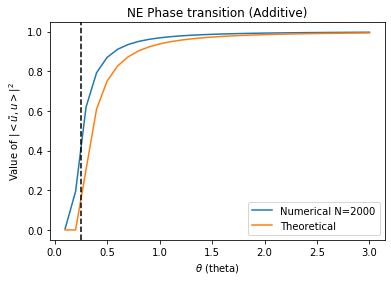}}
\caption{Comparison of transition in NE for additive perturbation of LQ game with $\sigma = 0.25$.}
\label{fig:4}
\end{figure}

To demonstrate the transition in NE, we use a numerical example with $n=2000$ Wigner matrix for additive perturbation as shown in \eqref{eq:7}. The value of $\theta$ is varied to see the effect of the threshold value of $\sigma$ on the Nash equilibrium. Since the NE of our LQ game is the leading eigenvector, we use the product in \eqref{eq:9} as the theoretical value and compare it with the computational result.
The vertical dashed line in Figure \ref{fig:3} and \ref{fig:4} is the critical transition point, across which the strength of the perturbation starts affecting the Nash equilibrium of the game. 

In Figure \ref{fig:3}, the critical threshold value of $\sigma = 1$. The value of $|<\tilde{u},u>|^2$ from \eqref{eq:9} is denoted by the 'theoretical' curve. The 'numerical' curve is derived by computationally calculating the NE, i.e the leading eigenvector for the perturbed adjacency matrix.

In Figure \ref{fig:4}, the critical threshold value of $\sigma = 0.25$. The numerical and the theoretical curves follow closely. The minor variation in the values of the two curves occurs due to the fact that \eqref{eq:9} is for range of $n \rightarrow \infty$. 

\section{Multiplicative Perturbation in LQ Game}
Now we explore another strategy to modify a network, through a multiplicative deformation. Assume $M_n$ be an $n \times m$ matrix with independent, zero mean, normally distributed entries with variance 1. Then, $X_n = M_nM_n^*/m $ is known as Wishart matrix. It has been shown that as $n,m \rightarrow \infty$ with $n/m \rightarrow c > 0$, the spectral measure of $X_n$ converges almost surely to the well-known Marchenko-Pastur distribution \cite{Marcenko1967} with density
\begin{equation} \label{eq:10}
    d\mu_x(x) = \frac{1}{2\pi cx} \sqrt{(b-x)(x-a)}1_{[a,b]}(x)dx + \max (0,1-\frac{1}{c}) \delta_0
\end{equation}
where $a=(1-\sqrt{c})^2$ and $b=(1+\sqrt{c})^2$. It is known that the extreme eigenvalues converge almost surely to the endpoints of this support.

Let perturbation matrix $P$, be a $n \times n$ symmetric matrix with rank r. It's eigenvalues are $\theta_1 \geq \dots \geq \theta_s >0 >\theta_{s+1} \geq \dots \geq \theta_r$. By phase transition theorem, as $n \rightarrow \infty$ we have, for 

\begin{equation} \label{eq:11}
\tilde{X} = X(I+P)
\end{equation}

\begin{equation} \label{eq:12}
    \lambda_{max}(\tilde{X}) \xrightarrow{\text{a.s.}} 
    \begin{cases}
    (\theta_1+1)(1 + \frac{c}{\theta_1}) & \text{if } \theta_1 \geq \sqrt{c} \\
    (1+\sqrt{c})^2 & \text{otherwise}
    \end{cases}
\end{equation}

In the setting where $P =\theta uu^{T}$, let $\tilde{u}$ be a unit-norm eigenvector of $\tilde{X}$ from \eqref{eq:11}, associated with its largest eigenvalue. For Wishart matrix $X$ as described above, the eigenvector phase transition occurs as,  

\begin{equation} \label{eq:13}
    |<\tilde{u},u>|^2 \xrightarrow{\text{a.s.}} 
    \begin{cases}
    \frac{\theta^2-c}{\theta[c(\theta+2)+\theta]} & \text{if } \theta \geq \sqrt{c} \\
    0 & \text{if } \theta < \sqrt{c}
    \end{cases}
\end{equation}

\subsection{Transitions in LQ game}
 In an LQ game, from \eqref{eq:1}, the large random matrix $X$ is considered to be $n \times n$ random Wishart matrix, with $P = \theta uu^T$ is the adjacency matrix of the perturbations so that $\theta$ and $u$ are largest eigenvalue and eigenvector respectively. Considering a multiplicative perturbation as in \eqref{eq:11}, the adjacency matrix of the new network is $\tilde{X}$. As seen in \eqref{eq:5}, the NE of $\tilde{X}$ is given by the eigenvector corresponding to maximum eigenvalue. Thus the NE $\tilde{x}^* = \lambda_{max}(\tilde{X}) = \tilde{u}$.

Using the result from \eqref{eq:13}, for the LQ game, the Nash Equilibrium of the perturbed game is dependent on the parameter $\theta$ of the perturbation.
A transition phenomenon is observed for values of $\theta$ greater than the threshold value of $\sqrt{c}$ when the NE given by $\tilde{u}$ is dependent on the network parameters i.e. $\theta$ and $\sqrt{c} $. Below this threshold value, the NE is not affected by the strength of the deformation and cannot be controlled by changing the edge weights of a network through external factors.

\subsection{Numerical Examples}
\begin{figure}[htbp]
\centerline{\includegraphics[width=0.99\columnwidth]{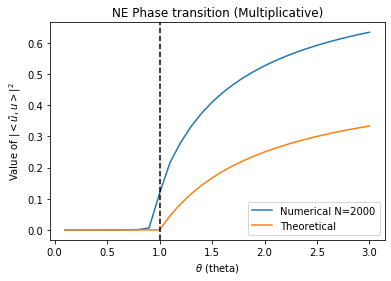}}
\caption{Comparison of transition in NE for multiplicative perturbation of LQ game with $m = 2000$, thus $\sqrt{c} = 1$.}
\label{fig:5}
\end{figure}

\begin{figure}[htbp]
\centerline{\includegraphics[width=0.99\columnwidth]{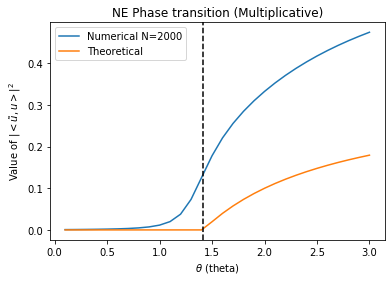}}
\caption{Comparison of transition in NE for multiplicative perturbation of LQ game with $m = 1000$, thus $\sqrt{c} = 1.414$.}
\label{fig:6}
\end{figure}

To demonstrate the transition of NE, we use a numerical example with $n=2000$ Wishart matrix for the multiplicative perturbation as shown in \eqref{eq:11}. The value of $\sqrt{c}$, dependent on $m$ as $n/m \rightarrow c$ is the threshold value given by dashed vertical line in Figure \ref{fig:5} and \ref{fig:6}. The value of $\theta$ is varied below and above $\sqrt{c}$ to observe the effect on the Nash equilibrium. Since the NE of our LQ game is the leading eigenvector, we use the product in \eqref{eq:13} as the theoretical value and compare it with the computational value.

In Figure \ref{fig:5}, the critical threshold value of $\sqrt{c} = 1$. The value of $|<\tilde{u},u>|^2$ from \eqref{eq:13} is denoted by the `theoretical' curve. The `numerical' curve is derived by computationally calculating the NE, i.e the leading eigenvector for the perturbed adjacency matrix. In Figure \ref{fig:6}, the critical threshold value of $\sqrt{c} = 1.414$. The numerical and the theoretical curve have differences which occur since in the numerical we have finite $n,m$ whereas for the `theoretical' we consider \eqref{eq:13} which represents the limit $n,m \rightarrow \infty$. In both examples, we observe that the NE depends on the strength of the deformation, when the strength is above the critical point there is phase transition for the NE.

\section{Interpretation}

In the previous sections, we see how a large network can be influenced through external perturbations. The Nash Equilibrium, for the LQ game in \eqref{eq:1}, can be affected through additive and multiplicative perturbation. If the perturbation strength $\theta$ is strong, the primary eigenvalue goes beyond the random spectrum and the primary eigenvector, which is the NE of our LQ game is correlated with $\theta$ (in a cone around the perturbation's eigenvector direction whose deviation angle goes to 0 as $\frac{\theta}{\sigma}\rightarrow \infty$). If $\theta$ is sufficiently low, the primary eigenvalue is buried in the random spectrum, and the NE is random, with no correlation to perturbation's eigenvector.

The finite rank perturbation of an adjacency matrix of large network can be interpreted in multiple ways. The additive perturbation can be designed such the specific nodes of the network has higher influence through their edges on the neighbors and consequently the entire network. The multiplicative perturbation, though harder to visualize, also affects the edge weights to dominate the influence from certain nodes and affect the NE properties. The NE, through eigenvalue phase transition is shown for rank 1 perturbation matrices. The rank 1 adjacency matrix can be designed is such a way that the influencers of a real world network have a major role to cause a aggressive spread over the network. The NE, which is the optimal participation of each member, is influenced by this perturbation. The shift in primary eigenvectors also affects the eigenvector centrality measure of the large networks.

The deformation to the networks can be manipulated such that influence of some of the important nodes of the network is powerful enough to propagate their beliefs quicker and effectively. Work in the area of identifying the subset of individuals within a network that can maximize spread of idea has studied like in \cite{Kempe2003MaximizingNetwork}. Using our analysis, we could apply sufficient external modifications to a network targeting influential set of individuals, to trigger a large cascade of belief spread.




\section{Conclusion}
In this paper, we explore conditions under which the Nash equilibrium of a large network game undergoes a phase transition when the adjacency matrix of the network is deformed through a finite matrix perturbation. Our contribution uses the well known eigenvector and eigenvalue phase transition phenomenon in conjunction with finite rank deformations of random matrices from field of random matrix theory to model such phenomena for linear quadratic games on a large network. For the LQ game, we make certain assumptions on the parameters of the utility function that enable us complete the analysis of transition of the NE. The large network is modelled as Winger and Wishart random matrix for additive and multiplicative deformations respectively. The network property i.e. $\sigma$ or $c$ determines the critical point around which  the strength of perturbation shifts the NE. We observe that, as $\theta$ crosses the critical point, the NE jumps out of the spectrum and is dependent on $\theta$. For the values of $\theta$ below the critical point, the NE is unaffected by $\theta$. There are multiple potential applications for this phenomenon, where selfish participants form a large network. 


There exists several avenues for future work. In field of network games, multiple settings are described through variety of utility functions. There is possibility of studying such phase transition in other network games with non linear functions. 
The eigenvalue/eigenvector transition for random matrix deformation occurs for distributions other than Wigner/Wishart as well. It would be a new direction to investigate phase transition for complex systems that are modeled by other density functions. Our model uses rank 1 deformation but studying phase transition for higher rank deformations and what such perturbations would mean to the network is another possible avenue. The extreme eigenvalues and spectral properties of adjacency matrix play an important role on other properties of NE like conditions on uniqueness and stability. Exploring how sudden shift in extreme eigenvalues affects these properties can help in devising appropriate perturbations. 
To this end, our results provide a framework to tackle some of the above open problems and applications in future work.

\bibliographystyle{ACM-Reference-Format}
\bibliography{references}


\begin{thebibliography}{30}


\ifx \showCODEN    \undefined \def \showCODEN     #1{\unskip}     \fi
\ifx \showDOI      \undefined \def \showDOI       #1{#1}\fi
\ifx \showISBNx    \undefined \def \showISBNx     #1{\unskip}     \fi
\ifx \showISBNxiii \undefined \def \showISBNxiii  #1{\unskip}     \fi
\ifx \showISSN     \undefined \def \showISSN      #1{\unskip}     \fi
\ifx \showLCCN     \undefined \def \showLCCN      #1{\unskip}     \fi
\ifx \shownote     \undefined \def \shownote      #1{#1}          \fi
\ifx \showarticletitle \undefined \def \showarticletitle #1{#1}   \fi
\ifx \showURL      \undefined \def \showURL       {\relax}        \fi
\providecommand\bibfield[2]{#2}
\providecommand\bibinfo[2]{#2}
\providecommand\natexlab[1]{#1}
\providecommand\showeprint[2][]{arXiv:#2}

\bibitem[\protect\citeauthoryear{Allouch}{Allouch}{2015}]%
        {Allouch2015}
\bibfield{author}{\bibinfo{person}{Nizar Allouch}.}
  \bibinfo{year}{2015}\natexlab{}.
\newblock \showarticletitle{{On the private provision of public goods on
  networks}}.
\newblock \bibinfo{journal}{\emph{Journal of Economic Theory}}
  (\bibinfo{year}{2015}).
\newblock
\showISSN{10957235}
\urldef\tempurl%
\url{https://doi.org/10.1016/j.jet.2015.01.007}
\showDOI{\tempurl}


\bibitem[\protect\citeauthoryear{Anderson, Guionnet, and Zeitouni}{Anderson
  et~al\mbox{.}}{2009}]%
        {Anderson2009}
\bibfield{author}{\bibinfo{person}{Greg~W. Anderson}, \bibinfo{person}{Alice
  Guionnet}, {and} \bibinfo{person}{Ofer Zeitouni}.}
  \bibinfo{year}{2009}\natexlab{}.
\newblock \bibinfo{booktitle}{\emph{{An Introduction to Random Matrices}}}.
\newblock \bibinfo{publisher}{Cambridge University Press}.
\newblock
\urldef\tempurl%
\url{https://doi.org/10.1017/cbo9780511801334}
\showDOI{\tempurl}


\bibitem[\protect\citeauthoryear{Baik, Arous, and Peche}{Baik
  et~al\mbox{.}}{2004}]%
        {Baik2004}
\bibfield{author}{\bibinfo{person}{Jinho Baik}, \bibinfo{person}{Gerard~Ben
  Arous}, {and} \bibinfo{person}{Sandrine Peche}.}
  \bibinfo{year}{2004}\natexlab{}.
\newblock \showarticletitle{{Phase transition of the largest eigenvalue for
  non-null complex sample covariance matrices}}.
\newblock \bibinfo{journal}{\emph{Annals of Probability}} \bibinfo{volume}{33},
  \bibinfo{number}{5} (\bibinfo{date}{3} \bibinfo{year}{2004}),
  \bibinfo{pages}{1643--1697}.
\newblock
\urldef\tempurl%
\url{http://arxiv.org/abs/math/0403022}
\showURL{%
\tempurl}


\bibitem[\protect\citeauthoryear{Baik and Silverstein}{Baik and
  Silverstein}{2006}]%
        {Baik2006}
\bibfield{author}{\bibinfo{person}{Jinho Baik} {and} \bibinfo{person}{Jack~W.
  Silverstein}.} \bibinfo{year}{2006}\natexlab{}.
\newblock \showarticletitle{{Eigenvalues of large sample covariance matrices of
  spiked population models}}.
\newblock \bibinfo{journal}{\emph{Journal of Multivariate Analysis}}
  \bibinfo{volume}{97}, \bibinfo{number}{6} (\bibinfo{date}{7}
  \bibinfo{year}{2006}), \bibinfo{pages}{1382--1408}.
\newblock
\showISSN{0047259X}
\urldef\tempurl%
\url{https://doi.org/10.1016/j.jmva.2005.08.003}
\showDOI{\tempurl}


\bibitem[\protect\citeauthoryear{Ballester, Calv{\'{o}}-Armengol, and
  Zenou}{Ballester et~al\mbox{.}}{2006}]%
        {Ballester2006}
\bibfield{author}{\bibinfo{person}{Coralio Ballester}, \bibinfo{person}{Antoni
  Calv{\'{o}}-Armengol}, {and} \bibinfo{person}{Yves Zenou}.}
  \bibinfo{year}{2006}\natexlab{}.
\newblock \showarticletitle{{Who's who in networks. Wanted: The key player}}.
\newblock \bibinfo{journal}{\emph{Econometrica}} (\bibinfo{year}{2006}).
\newblock
\showISSN{00129682}
\urldef\tempurl%
\url{https://doi.org/10.1111/j.1468-0262.2006.00709.x}
\showDOI{\tempurl}


\bibitem[\protect\citeauthoryear{Bassler, Forrester, and Frankel}{Bassler
  et~al\mbox{.}}{2008}]%
        {Bassler2008}
\bibfield{author}{\bibinfo{person}{Kevin~E. Bassler}, \bibinfo{person}{Peter~J.
  Forrester}, {and} \bibinfo{person}{Norman~E. Frankel}.}
  \bibinfo{year}{2008}\natexlab{}.
\newblock \showarticletitle{{Eigenvalue Separation in Some Random Matrix
  Models}}.
\newblock \bibinfo{journal}{\emph{J. Math. Phys.}} \bibinfo{volume}{50},
  \bibinfo{number}{3} (\bibinfo{date}{10} \bibinfo{year}{2008}).
\newblock
\urldef\tempurl%
\url{https://doi.org/10.1063/1.3081391}
\showDOI{\tempurl}


\bibitem[\protect\citeauthoryear{Benaych-Georges and
  Nadakuditi}{Benaych-Georges and Nadakuditi}{2009}]%
        {Benaych-Georges2009}
\bibfield{author}{\bibinfo{person}{Florent Benaych-Georges} {and}
  \bibinfo{person}{Raj~Rao Nadakuditi}.} \bibinfo{year}{2009}\natexlab{}.
\newblock \showarticletitle{{The eigenvalues and eigenvectors of finite, low
  rank perturbations of large random matrices}}.
\newblock  (\bibinfo{date}{10} \bibinfo{year}{2009}).
\newblock
\urldef\tempurl%
\url{http://arxiv.org/abs/0910.2120}
\showURL{%
\tempurl}


\bibitem[\protect\citeauthoryear{Bramoull{\'{e}}, Kranton, Bramoull{\'{e}}, and
  Kranton}{Bramoull{\'{e}} et~al\mbox{.}}{2016}]%
        {Bramoulle2016}
\bibfield{author}{\bibinfo{person}{Yann Bramoull{\'{e}}},
  \bibinfo{person}{Rachel Kranton}, \bibinfo{person}{Yann Bramoull{\'{e}}},
  {and} \bibinfo{person}{Rachel Kranton}.} \bibinfo{year}{2016}\natexlab{}.
\newblock \showarticletitle{{Games Played on Networks}}.
\newblock In \bibinfo{booktitle}{\emph{The Oxford Handbook of the Economics of
  Networks}}.
\newblock
\urldef\tempurl%
\url{https://doi.org/10.1093/oxfordhb/9780199948277.013.8}
\showDOI{\tempurl}


\bibitem[\protect\citeauthoryear{Calv{\'{o}}-Armengol, Patacchini, and
  Zenou}{Calv{\'{o}}-Armengol et~al\mbox{.}}{2009}]%
        {Calvo-Armengol2009}
\bibfield{author}{\bibinfo{person}{Antoni Calv{\'{o}}-Armengol},
  \bibinfo{person}{Eleonora Patacchini}, {and} \bibinfo{person}{Yves Zenou}.}
  \bibinfo{year}{2009}\natexlab{}.
\newblock \showarticletitle{{Peer effects and social networks in education}}.
\newblock \bibinfo{journal}{\emph{Review of Economic Studies}}
  (\bibinfo{year}{2009}).
\newblock
\showISSN{00346527}
\urldef\tempurl%
\url{https://doi.org/10.1111/j.1467-937X.2009.00550.x}
\showDOI{\tempurl}


\bibitem[\protect\citeauthoryear{Calv{\'{o}}-Armengol and
  Zenou}{Calv{\'{o}}-Armengol and Zenou}{2004}]%
        {Calvo-Armengol2004}
\bibfield{author}{\bibinfo{person}{Antoni Calv{\'{o}}-Armengol} {and}
  \bibinfo{person}{Yves Zenou}.} \bibinfo{year}{2004}\natexlab{}.
\newblock \bibinfo{title}{{Social networks and crime decisions: The role of
  social structure in facilitating delinquent behavior}}.
\newblock , \bibinfo{numpages}{939--958}~pages.
\newblock
\showISSN{00206598}
\urldef\tempurl%
\url{https://doi.org/10.1111/j.0020-6598.2004.00292.x}
\showDOI{\tempurl}


\bibitem[\protect\citeauthoryear{Capitaine, Donati-Martin, and
  F{\'{e}}ral}{Capitaine et~al\mbox{.}}{2009}]%
        {Capitaine2009}
\bibfield{author}{\bibinfo{person}{Mireille Capitaine},
  \bibinfo{person}{Catherine Donati-Martin}, {and} \bibinfo{person}{Delphine
  F{\'{e}}ral}.} \bibinfo{year}{2009}\natexlab{}.
\newblock \showarticletitle{{The largest eigenvalues of finite rank deformation
  of large wigner matrices: Convergence and nonuniversality of the
  fluctuations}}.
\newblock \bibinfo{journal}{\emph{Annals of Probability}} \bibinfo{volume}{37},
  \bibinfo{number}{1} (\bibinfo{date}{1} \bibinfo{year}{2009}),
  \bibinfo{pages}{1--47}.
\newblock
\showISSN{00911798}
\urldef\tempurl%
\url{https://doi.org/10.1214/08-AOP394}
\showDOI{\tempurl}


\bibitem[\protect\citeauthoryear{F{\'{e}}ral and P{\'{e}}ch{\'{e}}}{F{\'{e}}ral
  and P{\'{e}}ch{\'{e}}}{2007}]%
        {Feral2007}
\bibfield{author}{\bibinfo{person}{Delphine F{\'{e}}ral} {and}
  \bibinfo{person}{Sandrine P{\'{e}}ch{\'{e}}}.}
  \bibinfo{year}{2007}\natexlab{}.
\newblock \showarticletitle{{The largest eigenvalue of rank one deformation of
  large wigner matrices}}.
\newblock \bibinfo{journal}{\emph{Communications in Mathematical Physics}}
  \bibinfo{volume}{272}, \bibinfo{number}{1} (\bibinfo{date}{5}
  \bibinfo{year}{2007}), \bibinfo{pages}{185--228}.
\newblock
\showISSN{00103616}
\urldef\tempurl%
\url{https://doi.org/10.1007/s00220-007-0209-3}
\showDOI{\tempurl}


\bibitem[\protect\citeauthoryear{Gladwell}{Gladwell}{2002}]%
        {Gladwell2002}
\bibfield{author}{\bibinfo{person}{Malcolm Gladwell}.}
  \bibinfo{year}{2002}\natexlab{}.
\newblock \bibinfo{booktitle}{\emph{{The Tipping Point: How Little Things Can
  Make a Big Difference.}}}
\newblock \bibinfo{publisher}{Little, Brown and Co}.
\newblock


\bibitem[\protect\citeauthoryear{Hoyle and Rattray}{Hoyle and Rattray}{2007}]%
        {Hoyle2007}
\bibfield{author}{\bibinfo{person}{D.~C. Hoyle} {and} \bibinfo{person}{M.
  Rattray}.} \bibinfo{year}{2007}\natexlab{}.
\newblock \showarticletitle{{Statistical mechanics of learning multiple
  orthogonal signals: Asymptotic theory and fluctuation effects}}.
\newblock \bibinfo{journal}{\emph{Physical Review E - Statistical, Nonlinear,
  and Soft Matter Physics}} \bibinfo{volume}{75}, \bibinfo{number}{1}
  (\bibinfo{year}{2007}).
\newblock
\showISSN{15502376}
\urldef\tempurl%
\url{https://doi.org/10.1103/PhysRevE.75.016101}
\showDOI{\tempurl}


\bibitem[\protect\citeauthoryear{Jackson}{Jackson}{2010}]%
        {Jackson2010}
\bibfield{author}{\bibinfo{person}{Matthew~O Jackson}.}
  \bibinfo{year}{2010}\natexlab{}.
\newblock \bibinfo{booktitle}{\emph{{Social and Economic Networks}}}.
\newblock \bibinfo{publisher}{Princeton university press}.
\newblock


\bibitem[\protect\citeauthoryear{Jackson and Zenou}{Jackson and Zenou}{2015}]%
        {Jackson2015}
\bibfield{author}{\bibinfo{person}{Matthew~O. Jackson} {and}
  \bibinfo{person}{Yves Zenou}.} \bibinfo{year}{2015}\natexlab{}.
\newblock \showarticletitle{{Games on Networks}}.
\newblock In \bibinfo{booktitle}{\emph{Handbook of Game Theory with Economic
  Applications}}.
\newblock
\showISSN{15740005}
\urldef\tempurl%
\url{https://doi.org/10.1016/B978-0-444-53766-9.00003-3}
\showDOI{\tempurl}


\bibitem[\protect\citeauthoryear{Karoui}{Karoui}{2005}]%
        {Karoui2005}
\bibfield{author}{\bibinfo{person}{Noureddine~El Karoui}.}
  \bibinfo{year}{2005}\natexlab{}.
\newblock \showarticletitle{{Tracy--Widom limit for the largest eigenvalue of a
  large class of complex sample covariance matrices}}.
\newblock \bibinfo{journal}{\emph{Annals of Probability}} \bibinfo{volume}{35},
  \bibinfo{number}{2} (\bibinfo{date}{3} \bibinfo{year}{2005}),
  \bibinfo{pages}{663--714}.
\newblock
\urldef\tempurl%
\url{https://doi.org/10.1214/009117906000000917}
\showDOI{\tempurl}


\bibitem[\protect\citeauthoryear{Kempe, Kleinberg, and Tardos}{Kempe
  et~al\mbox{.}}{2003}]%
        {Kempe2003MaximizingNetwork}
\bibfield{author}{\bibinfo{person}{David Kempe}, \bibinfo{person}{Jon
  Kleinberg}, {and} \bibinfo{person}{Éva Tardos}.}
  \bibinfo{year}{2003}\natexlab{}.
\newblock \showarticletitle{{Maximizing the spread of influence through a
  social network}}. In \bibinfo{booktitle}{\emph{Proceedings of the ACM SIGKDD
  International Conference on Knowledge Discovery and Data Mining}}.
  \bibinfo{pages}{137--146}.
\newblock
\urldef\tempurl%
\url{https://doi.org/10.1145/956750.956769}
\showDOI{\tempurl}


\bibitem[\protect\citeauthoryear{Koenig, Liu, and Zenou}{Koenig
  et~al\mbox{.}}{2014}]%
        {Koenig2014}
\bibfield{author}{\bibinfo{person}{Michael~D. Koenig},
  \bibinfo{person}{Xiaodong Liu}, {and} \bibinfo{person}{Yves Zenou}.}
  \bibinfo{year}{2014}\natexlab{}.
\newblock \showarticletitle{{R{\&}D Networks: Theory, Empirics and Policy
  Implications}}.
\newblock \bibinfo{journal}{\emph{Discussion Papers}} (\bibinfo{year}{2014}).
\newblock
\urldef\tempurl%
\url{https://ideas.repec.org/p/sip/dpaper/13-027.html}
\showURL{%
\tempurl}


\bibitem[\protect\citeauthoryear{Landau}{Landau}{1976}]%
        {Landau1976Finite-sizeLattice}
\bibfield{author}{\bibinfo{person}{D.~P. Landau}.}
  \bibinfo{year}{1976}\natexlab{}.
\newblock \showarticletitle{{Finite-size behavior of the Ising square
  lattice}}.
\newblock \bibinfo{journal}{\emph{Physical Review B}} \bibinfo{volume}{13},
  \bibinfo{number}{7} (\bibinfo{date}{4} \bibinfo{year}{1976}),
  \bibinfo{pages}{2997--3011}.
\newblock
\showISSN{01631829}
\urldef\tempurl%
\url{https://doi.org/10.1103/PhysRevB.13.2997}
\showDOI{\tempurl}


\bibitem[\protect\citeauthoryear{Mar{\v{c}}enko and Pastur}{Mar{\v{c}}enko and
  Pastur}{1967}]%
        {Marcenko1967}
\bibfield{author}{\bibinfo{person}{V~A Mar{\v{c}}enko} {and}
  \bibinfo{person}{L~A Pastur}.} \bibinfo{year}{1967}\natexlab{}.
\newblock \showarticletitle{{DISTRIBUTION OF EIGENVALUES FOR SOME SETS OF
  RANDOM MATRICES}}.
\newblock \bibinfo{journal}{\emph{Mathematics of the USSR-Sbornik}}
  \bibinfo{volume}{1}, \bibinfo{number}{4} (\bibinfo{date}{4}
  \bibinfo{year}{1967}), \bibinfo{pages}{457--483}.
\newblock
\showISSN{0025-5734}
\urldef\tempurl%
\url{https://doi.org/10.1070/sm1967v001n04abeh001994}
\showDOI{\tempurl}


\bibitem[\protect\citeauthoryear{McLeman, Dupre, Berrang~Ford, Ford, Gajewski,
  and Marchildon}{McLeman et~al\mbox{.}}{2014}]%
        {McLeman2014}
\bibfield{author}{\bibinfo{person}{Robert~A. McLeman},
  \bibinfo{person}{Juliette Dupre}, \bibinfo{person}{Lea Berrang~Ford},
  \bibinfo{person}{James Ford}, \bibinfo{person}{Konrad Gajewski}, {and}
  \bibinfo{person}{Gregory Marchildon}.} \bibinfo{year}{2014}\natexlab{}.
\newblock \showarticletitle{{What we learned from the Dust Bowl: lessons in
  science, policy, and adaptation}}.
\newblock \bibinfo{journal}{\emph{Population and Environment}}
  \bibinfo{volume}{35}, \bibinfo{number}{4} (\bibinfo{date}{6}
  \bibinfo{year}{2014}), \bibinfo{pages}{417--440}.
\newblock
\showISSN{0199-0039}
\urldef\tempurl%
\url{https://doi.org/10.1007/s11111-013-0190-z}
\showDOI{\tempurl}


\bibitem[\protect\citeauthoryear{Meyers and Rhode}{Meyers and Rhode}{2019}]%
        {Meyers2019}
\bibfield{author}{\bibinfo{person}{Keith Meyers} {and} \bibinfo{person}{Paul~W
  Rhode}.} \bibinfo{year}{2019}\natexlab{}.
\newblock \bibinfo{booktitle}{\emph{{Exploring the Causes Driving Hybrid Corn
  Adoption from 1933 to 1955}}}.
\newblock \bibinfo{type}{{T}echnical {R}eport}.
\newblock


\bibitem[\protect\citeauthoryear{Moscona, Acemoglu, Atkin, Cohen, Krieger,
  Lerner, Moser, Nunn, Olken, Sastry, and Shleifer}{Moscona
  et~al\mbox{.}}{2021}]%
        {Moscona2021}
\bibfield{author}{\bibinfo{person}{Jacob Moscona}, \bibinfo{person}{Daron
  Acemoglu}, \bibinfo{person}{David Atkin}, \bibinfo{person}{Lauren Cohen},
  \bibinfo{person}{Joshua~Lev Krieger}, \bibinfo{person}{Josh Lerner},
  \bibinfo{person}{Petra Moser}, \bibinfo{person}{Nathan Nunn},
  \bibinfo{person}{Ben Olken}, \bibinfo{person}{Karthik Sastry}, {and}
  \bibinfo{person}{Andrei Shleifer}.} \bibinfo{year}{2021}\natexlab{}.
\newblock \bibinfo{booktitle}{\emph{{Environmental Catastrophe and the
  Direction of Invention: Evidence from the American Dust Bowl *}}}.
\newblock \bibinfo{type}{{T}echnical {R}eport}.
\newblock
\urldef\tempurl%
\url{http://economics.mit.edu/grad/moscona}
\showURL{%
\tempurl}


\bibitem[\protect\citeauthoryear{Nadler}{Nadler}{2008}]%
        {Nadler2008}
\bibfield{author}{\bibinfo{person}{Boaz Nadler}.}
  \bibinfo{year}{2008}\natexlab{}.
\newblock \showarticletitle{{Finite sample approximation results for principal
  component analysis: A matrix perturbation approach}}.
\newblock \bibinfo{journal}{\emph{Annals of Statistics}} \bibinfo{volume}{36},
  \bibinfo{number}{6} (\bibinfo{date}{12} \bibinfo{year}{2008}),
  \bibinfo{pages}{2791--2817}.
\newblock
\showISSN{00905364}
\urldef\tempurl%
\url{https://doi.org/10.1214/08-AOS618}
\showDOI{\tempurl}


\bibitem[\protect\citeauthoryear{Naghizadeh and Liu}{Naghizadeh and
  Liu}{2017}]%
        {Naghizadeh2017}
\bibfield{author}{\bibinfo{person}{Parinaz Naghizadeh} {and}
  \bibinfo{person}{Mingyan Liu}.} \bibinfo{year}{2017}\natexlab{}.
\newblock \showarticletitle{{On the uniqueness and stability of equilibria of
  network games}}.
\newblock \bibinfo{journal}{\emph{55th Annual Allerton Conference on
  Communication, Control, and Computing, Allerton 2017}}
  \bibinfo{volume}{2018-Janua} (\bibinfo{year}{2017}),
  \bibinfo{pages}{280--286}.
\newblock
\showISBNx{9781538632666}
\urldef\tempurl%
\url{https://doi.org/10.1109/ALLERTON.2017.8262749}
\showDOI{\tempurl}


\bibitem[\protect\citeauthoryear{Naghizadeh and Liu}{Naghizadeh and
  Liu}{2018}]%
        {Naghizadeh2018}
\bibfield{author}{\bibinfo{person}{Parinaz Naghizadeh} {and}
  \bibinfo{person}{Mingyan Liu}.} \bibinfo{year}{2018}\natexlab{}.
\newblock \showarticletitle{{Provision of Public Goods on Networks: On
  Existence, Uniqueness, and Centralities}}.
\newblock \bibinfo{journal}{\emph{IEEE Transactions on Network Science and
  Engineering}} (\bibinfo{year}{2018}).
\newblock
\showISSN{23274697}
\urldef\tempurl%
\url{https://doi.org/10.1109/TNSE.2017.2755003}
\showDOI{\tempurl}


\bibitem[\protect\citeauthoryear{Newman}{Newman}{2018}]%
        {Newman2018}
\bibfield{author}{\bibinfo{person}{Mark Newman}.}
  \bibinfo{year}{2018}\natexlab{}.
\newblock \bibinfo{booktitle}{\emph{{Networks}}}.
\newblock \bibinfo{publisher}{Oxford university press}.
\newblock


\bibitem[\protect\citeauthoryear{Parise and Ozdaglar}{Parise and
  Ozdaglar}{2017}]%
        {Parise2017a}
\bibfield{author}{\bibinfo{person}{Francesca Parise} {and}
  \bibinfo{person}{Asuman Ozdaglar}.} \bibinfo{year}{2017}\natexlab{}.
\newblock \showarticletitle{{A variational inequality framework for network
  games: Existence, uniqueness, convergence and sensitivity analysis}}.
\newblock \bibinfo{journal}{\emph{Games and Economic Behavior}}
  \bibinfo{volume}{114} (\bibinfo{date}{12} \bibinfo{year}{2017}),
  \bibinfo{pages}{47--82}.
\newblock
\urldef\tempurl%
\url{http://arxiv.org/abs/1712.08277}
\showURL{%
\tempurl}


\bibitem[\protect\citeauthoryear{Stanley}{Stanley}{1999}]%
        {Stanley1999ScalingPhenomena}
\bibfield{author}{\bibinfo{person}{H.~Eugene Stanley}.}
  \bibinfo{year}{1999}\natexlab{}.
\newblock \bibinfo{title}{{Scaling, universality, and renormalization: Three
  pillars of modern critical phenomena}}.
\newblock , \bibinfo{numpages}{S358}~pages.
\newblock
\showISSN{00346861}
\urldef\tempurl%
\url{https://doi.org/10.1103/revmodphys.71.s358}
\showDOI{\tempurl}


\end{thebibliography}

\end{document}